 \documentclass[preprint,journal]{vgtc}       

\ifpdf
  \pdfoutput=1\relax                   
  \usepackage{graphicx}                
  \DeclareGraphicsExtensions{.pdf,.png,.jpg,.jpeg} 
\else
  \usepackage{graphicx}                
  \DeclareGraphicsExtensions{.eps}     
\fi%

\graphicspath{{figures/}{pictures/}{images/}{./}} 

\usepackage{microtype}                 
\PassOptionsToPackage{warn}{textcomp}  
\usepackage{textcomp}                  
\usepackage{mathptmx}                  
\usepackage{times}                     
\usepackage{cite}                      
\usepackage{tabu}                      
\usepackage{booktabs}                  

\usepackage{url}

\usepackage[draft]{todonotes} 
\usepackage{scalerel}

\usepackage{subfigure}

\usepackage{xcolor,colortbl}
\usepackage{makecell}
\usepackage{multirow}

\newcommand*{\SmallIndent}{\hspace*{0.27cm}}
\usepackage{soul}

\colorlet{firstround}{black}
\colorlet{luke}{black}
\colorlet{calvin}{black}
\colorlet{luke2}{black}

\ieeedoi{10.1109/TVCG.2019.2934603}

\onlineid{1260}

\vgtccategory{Research}
\vgtcpapertype{application/design study}

\title{\textcolor{firstround}{MetricsVis: A Visual Analytics System for Evaluating Employee Performance in Public Safety Agencies}}

\author{Jieqiong~Zhao, Morteza~Karimzadeh, Luke~S.~Snyder, Chittayong~Surakitbanharn,\\Zhenyu~Cheryl~Qian, and~David~S.~Ebert,~\textit{Fellow,~IEEE}}
\authorfooter{
\item Jieqiong Zhao, Luke S. Snyder, Zhenyu Cheryl Qian, and David S. Ebert are with Purdue University.\\
E-mail: \{zhao413, snyde238, qianz, ebertd\}@purdue.edu
\item Morteza Karimzadeh is with University of Colorado Boulder (formerly at Purdue University). E-mail: \{karimzadeh\}@colorado.edu
\item Chittayong Surakitbanharn is with Hivemapper. \\
E-mail: \{chittayong.surakitbanharn\}@gmail.com
}

\shortauthortitle{Zhao \MakeLowercase{\textit{et al.}}: MetricsVis: A Visual Analytics System for Evaluating Employee Performance in Public Safety Agencies}

\abstract{Evaluating employee performance in organizations with varying workloads and tasks is challenging. 
Specifically, it is important to understand how quantitative measurements of employee achievements relate to supervisor expectations, what the main drivers of good performance are, and how to combine these complex and flexible performance evaluation metrics into an accurate portrayal of organizational performance in order to  
identify shortcomings and improve overall productivity. 
To facilitate this process, we summarize common organizational performance analyses into four visual exploration task categories.
Additionally, we develop MetricsVis, a visual analytics \textcolor{firstround}{system} composed of multiple coordinated views to support the dynamic evaluation and comparison of individual, team, and organizational performance \textcolor{firstround}{in public safety organizations}.
\textcolor{firstround}{MetricsVis provides four primary visual components to expedite performance evaluation:}
(1) a priority adjustment view to support direct manipulation on evaluation metrics;
(2) a reorderable performance matrix to demonstrate the details of individual employees;
(3) a group performance view that highlights aggregate performance and individual contributions for each group;
and (4) a projection view illustrating employees with similar specialties to facilitate shift assignments and training.
We demonstrate the usability of our framework with two case studies from medium-sized law enforcement agencies and highlight  its broader applicability to other domains.
} 


\CCScatlist{ 
 \CCScat{K.6.1}{Management of Computing and Information Systems}%
{Project and People Management}{Life Cycle};
 \CCScat{K.7.m}{The Computing Profession}{Miscellaneous}{Ethics}
}

\teaser{
  \centering
  \includegraphics[width=\linewidth]{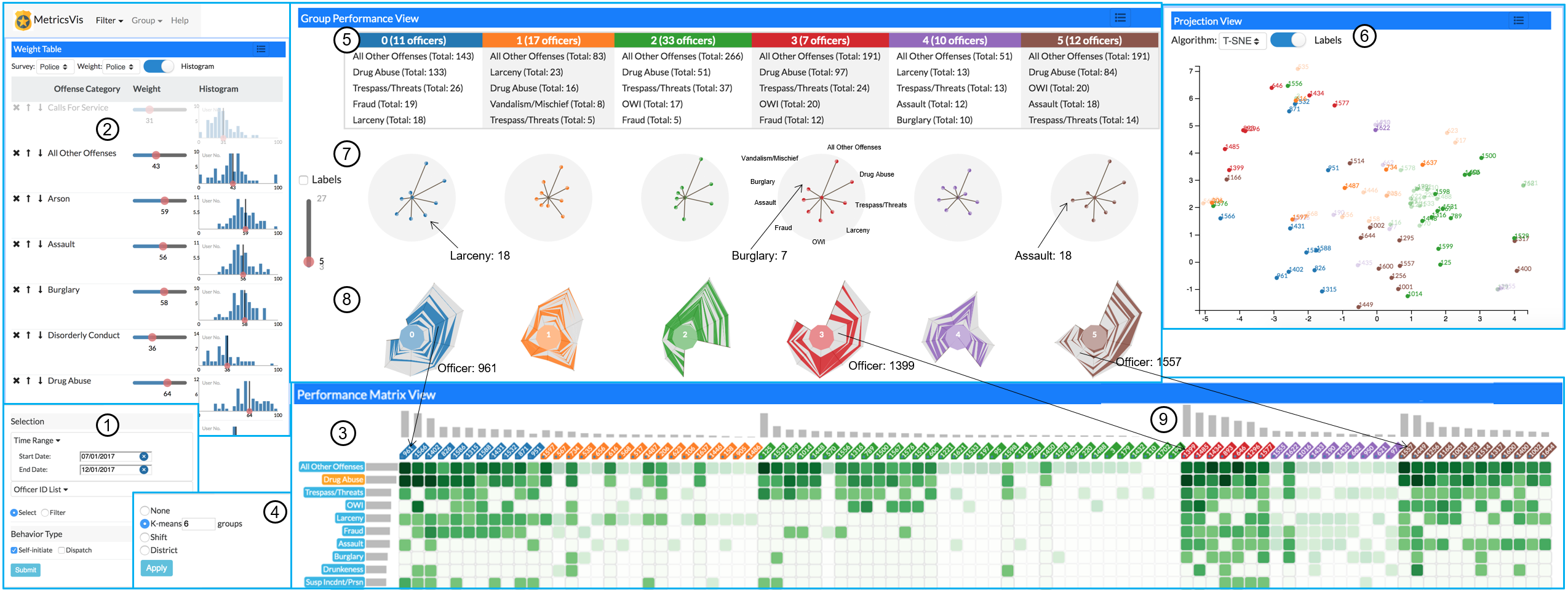}
  \caption{MetricsVis overview: 
    The \textcolor{firstround}{priority adjustment view} (2) encodes the crowdsourced crime severity ratings from police officers and citizens (perceived importance of factors); the red dots indicate the currently assigned weights used in the evaluation metrics.
    The projection view (6) shows the dimensionality reduction results. 
    The group performance view (5) contains three visual representations that show an overview of group performance and the contribution of each member.
    The performance matrix view (3) displays the individual employee performance with employees in columns and job types in rows (here, employees are sorted based on their group first and then their total performance scores).
    The control panel shows the filters (1) and grouping method (4) applied in use case 1.
  }
  \label{fig:teaser}
}

\vgtcinsertpkg

\begin{document}


\firstsection{Introduction}

\maketitle


Organizational performance evaluation can be useful for strategic planning, staff management, and operational development.
An effective performance evaluation system with clearly defined goals and prompt feedback is essential for organizations to improve their productivity~\cite{Latham1981increasing}, especially with limited resources and personnel.
Oftentimes, employee, unit, and organizational performance characterization requires considering multiple facets such as economic return, social impact, sustainability, and team and individual productivity.
However, developing an appropriate method to integrate multiple aspects, including qualitative, quantitative, and subjective data into an accurate representation of an organization's \textcolor{firstround}{performance} is challenging.
The task is further complicated when different employee teams have different workloads based on shifts and locations, \textcolor{firstround}{as is common in public safety organizations}.

To address this problem, it can be beneficial to interactively analyze, visually explore, accurately weight, compare, and evaluate \textcolor{firstround}{employee performance in the context of organizational hierarchy}. 
Applying standardized evaluation factors and quantitative metrics can help overcome stereotypical biases caused by personal traits~\cite{smither1998lessons,bernardin1984performance}.
Such metrics should account for the importance of each task type in reaching the objectives of an organization, while accommodating the perspectives of team leaders across various departments.

Currently, visual analytics tools specialized for harnessing multi-dimensional organizational data to facilitate effective employee performance evaluation  are lacking.
Current performance evaluation practices often apply subjective supervisory ratings with data tables listing the simple statistical summaries of entire departments and details regarding tasks completed by individual employees. 
Existing visual analytics applications support either multi-dimensional data visualizations or multi-criteria decision-making that treat individuals uniformly,  ignoring the inherent hierarchical relationships, different teams or task types typical to \textcolor{firstround}{public safety and many other} organizations\textcolor{luke}{~\cite{zhao2017skylens, gratzl2013lineup}}.

In this \textcolor{luke}{paper}, we present MetricsVis (Fig.~\ref{fig:teaser}), a visual analytics \textcolor{firstround}{system} for evaluating the performance of individual employees, teams, and the entire organization \textcolor{luke}{in public safety agencies}.
We designed the \textcolor{firstround}{system} iteratively with users from two medium-sized law enforcement agencies (representing similar-sized organizations in our study).
We rooted our metrics in the existing organizational performance literature and adaptively tailored MetricsVis to meet the requirements of \textcolor{firstround}{public safety} organizations with groups of employees performing similar jobs but at different locations and times, resulting in different workloads that impact their contribution to organizational goals.
Additionally, we formalized the analytical tasks, goals, and metrics; derived metrics; and surveyed organizational personnel and the public to decide on priorities of evaluation.
We implemented multiple coordinated views in MetricsVis to support efficient, effective, and dynamic performance evaluation for multiple levels of an organization. 

The MetricsVis \textcolor{firstround}{system} enables a holistic evaluation of organizational priorities versus actual achievements, and helps identify opportunities for improvement.
Additionally, it facilitates evaluation of strategic goals, expedites resource allocation (e.g. understanding which employees may need additional training or would be good trainers), and improves workload balance and individual employee performance.
Specific contributions of our research and design are as follows:
\begin{itemize}
\item The mapping of the analysis of public-safety organizational performance evaluation into four visual analytical task categories. 
\item A novel \textcolor{firstround}{system} supporting interactive visual organizational performance analysis \textcolor{luke}{in public safety agencies} based on hybrid evaluation metrics that integrate quantitative employee data and qualitative subjective feedback, and appropriate visual representations to support the four aforementioned visual task categories.
\item A \textcolor{firstround}{system} evaluation   with domain experts from two medium-sized law enforcement agencies to validate system usability.
\end{itemize}

\section{Related Work}\label{sec:related-work}

\textcolor{luke}{In this section, we review previous work on performance evaluation in organizations, interactive sorting techniques that aid in effective ranking and comparison of multi-dimensional data, and analytical systems to make sense of multi-attribute data.}

\subsection{Performance Evaluation in Organizations}

Performance evaluation needs to compare performance observations with expectations, reveal barriers preventing the desired performance, and generate action plans for either maintenance or improvement to achieve organizational objectives~\cite{guerra2007evaluating,guerra2008performance}.
Performance appraisal systems assist in realigning employee performance to meet the evolving organizational objectives~\cite{coutts2004police}.
An ongoing problem in organizational performance is designing metrics to measure employee effectiveness and productivity~\cite{kanter1981organizational,becker1996impact}.

Several researchers have derived taxonomies to evaluate employee performance-related factors that characterize the performance of individual employees~\cite{borman1993more,arvey1998performance,koopmans2011conceptual, cascio2011applied, campbell1993performance, Borman1993expanding, sackett2002counterproductive}.
The results are lists of generalizable evaluation factors 
(e.g. task performance, organizational citizenship behavior, and counterproductive work behavior)
that could be adopted in diverse evaluation scenarios.
MetricsVis leverages dynamic evaluation factors which users can customize based on organizational objectives, and supports interactive variable weighting to reflect the relative importance of each task/job type (i.e., factor).
In addition to evaluating individual performance, the hierarchical structure in an organization has a fundamental impact on the organization's behavior and management~\cite{ivancevich1999organizational,delaney1996impact}.
Our \textcolor{firstround}{system} supports the exploration of individual performance as well as team- and organization-level performance with respect to the organizational hierarchy.


\subsection{Interactive Sorting and Visualization}

Tabular visualizations have been widely used~\cite{rao1994table} and adapted to support hierarchical structure models (e.g. ValueCharts~\cite{carenini2004valuecharts}), multi-attribute ranking systems (e.g., LineUp ~\cite{gratzl2013lineup}) , and interactive categorical exploration (e.g., parallel sets~\cite{kosara2006parallelset}).
Specifically, reorderable matrices efficiently explore associations between hundreds of data items and data attributes~\cite{mclachlan2008liverac}.
Furthermore, permutation (sorting, clustering) methods help highlight similarity patterns in these matrices ~\cite{behrisch2016matrix}.

Interactive ranking and sorting is an active  research area. 
Some techniques sort all data attributes simultaneously and use linkages across all attributes to highlight the same data entry~\cite{xia2017rank, hur2009simulsort}. Timespan~\cite{loorak2016timespan} supports hierarchical reordering, which sorts data samples based on the priority of a data attribute.
However, MetricsVis provides conventional sorting on a reorderable matrix that allows flexible rearranging of attributes (i.e., factors such as job categories) and data items (employees), because we found that these components are more familiar to end users.

MetricsVis computes ranking based on attribute weights provided by end users, as opposed to several recent systems that leverage user-assigned data ordering to reverse engineer the weights ~\cite{wall2017podium, zanakis1998madm}.
These systems require users to interactively update the overall ranking of data samples and inspect the validity of weights.
MetricsVis, however, requires domain experts to have a good understanding of the weights, which is coherent with the goal of aligning with the priorities of an organization for dynamic evaluation purpose.

\subsection{\textcolor{luke}{Visual Analytics for Multi-Attribute Decision Making}}
\textcolor{luke}{
Researchers have presented several visual analytic (VA) systems to facilitate the exploration and understanding of multi-dimensional data.
Zhao et al.~\cite{zhao2017skylens} developed SkyLens, a VA solution that enables comparison of multi-dimensional data through multiple coordinated views, while filtering out inferior data candidates.
LineUp~\cite{gratzl2013lineup}, perhaps the work most similar to ours, performs ranking visualization of multi-attribute data, and allows users to flexibly adjust weighting parameters to identify potential relationships.
However, SkyLens and LineUp do not provide interactive visualizations to support multi-level performance comparison, such as
individual to group or group to organization, which is necessary for organizational evaluation. MetricsVis was designed with this key consideration in mind. In addition, LineUp utilizes bar charts to facilitate ranking comparison; MetricsVis employs radial layouts, which have outperformed tabular layouts when comparing data attributes~\cite{keck2017glyph} and provide compact visualization.}

\textcolor{luke}{
The software suite Tableau~\cite{Tableau} can provide useful individual interactive data visualizations to explore relationships, trends, and rankings among multi-attribute data, such as pie, bubble, bar charts, treemaps, and tabular visuals.
However, the user may not be able to generate the visualization that communicates the data most effectively to compare multi-level performance.
MetricsVis provides compact and interactively linked visualizations specifically tailored for efficient, multi-level comparison of organizational performance metrics.
For instance, MetricsVis allows users to view individuals with potentially similar performance through integrated clustering. Tableau does not support this.
Furthermore, while Tableau can provide a hierarchical overview of an individual's contribution to the group (and group to organization) with treemap visualization, MetricsVis supports simultaneous comparison of multiple attributes to the overall group with stacked radar charts, which can also be used to compare between groups.
}



\section{\textcolor{firstround}{Domain Characterization}}\label{sec:task}
\textcolor{firstround}{We identified the general requirements for an effective and efficient performance evaluation system by reviewing the literature~\cite{guerra2007evaluating, guerra2008performance, coutts2004police, kanter1981organizational, becker1996impact, ivancevich1999organizational, delaney1996impact, cascio2011applied, Sanchez2011risk, campion2011competencies}, which informed our discussions  with domain experts from law enforcement. We then mapped the refined requirements into four visual analytical task categories, as explained below.}

\subsection{Requirements Analysis}

Assessing employee performance involves considering and integrating multiple performance-contributing factors to enable accurate comparison against organizational objectives.
To satisfy the diverse requirements of different organizations, we decided to design on-demand comprehensive evaluation metrics, so that users can dynamically choose their preferred metrics.
For clarification, evaluation metrics in our context are comprised of two aspects: the performance-contributing factors and their weights.
Choosing performance factors is a challenging task specific to each domain.
We derived these factors from a combination of (a) unstructured interviews with commanders and chiefs at law enforcement agencies and (b) taxonomies of \textcolor{firstround}{task} performance in work settings~\cite{borman1993more,arvey1998performance,koopmans2011conceptual, cascio2011applied, Sanchez2011risk, campion2011competencies}. 

Organizational hierarchy affects the performance evaluation process.
Evaluators' differing perspectives can hinder comparisons across the entire department, especially when traditional performance evaluation practices rely heavily on subjective ratings from management.
Leaders may analyze their team's workload and performance quality through personal interaction and job activity reports. 
However, it may still be difficult to compare different aspects of satisfactory performance between units across the organization.
Though the evaluators can ensure unbiased judgment within their teams, variation across multiple evaluators is inevitable.
Besides comparisons at the same level (individual versus individual, group versus group), we also need to evaluate the contribution of an individual to its group and a group to its organization.
Therefore, an effective performance management system must support the performance evaluation at and between multiple levels of the organizational hierarchy.

We summarize these requirements from three independent perspectives
: (1) dynamic performance metrics that can be adjusted by users to align with organizational objectives; (2) multiple \textcolor{firstround}{levels} including individual, team, and the entire organization; and (3) two relational contexts including comparisons at the same level as well as between two levels.


\subsection{Analytical Tasks}\label{subsec:analytical_tasks}

The goal of MetricsVis is to enable the evaluation of individual employee, team, and organization effectiveness through the exploration of performance measures derived from digitized data (quantitative, qualitative, and subjective).
To accomplish this goal, MetricsVis was designed to address several visual analytical task categories for performance evaluation: 

\begin{enumerate}
\item [\textbf{T1}] \textbf{Evaluate individual employee performance:} 
The first challenge a team manager may encounter is aggregating and transforming activity reports and statistics into measures of subordinate performance. 
One approach to evaluating performance is determining the frequency of different jobs accomplished by an employee, the difficulty and effort required for a certain category of job,  and whether the job was self-initiated or dispatched. 
Since not all job types are equal in difficulty and effort, the option to weight each job type should be incorporated.
Additionally, a supervisor needs to select and compare multiple employees' performance to find patterns in low- and high-performing employees.

\item [\textbf{T2}] \textbf{Evaluate group and team performance:} 
The second challenge is understanding the most influential factors that create successful teams for a variety of jobs. 
Some factors that may impact team effectiveness include location, manager, shift time of day, personnel proficiency levels, and time spent working.
For instance, assigning police officers with experience in a certain geographic area to respond to calls in that area may increase patrolling effectiveness due to additional tacit knowledge providing an advantage over someone unfamiliar with that area. 
Additionally, understanding how these issues affect workload balance and morale serves to help optimize personnel allocation. 

\item [\textbf{T3}] \textbf{Investigate organizational workload:}
Managers want to understand resource and personnel allocation strategies, pattern changes in services, and the effect on workload balance to increase overall organizational effectiveness.
Exploring and comparing grouping factors (e.g. locations, time periods, and servicing patterns) can enable understanding of whether resource expenditure is aligned with organizational goals, discover unexpected drains on resources, and find excess personnel capacity in certain areas or during certain time periods.
This information is crucial for advanced resource allocation strategies (e.g. dynamic allocation, request-based allocation), and evaluating the effectiveness of alternative strategies. 

\item [\textbf{T4}] \textbf{Evaluate department priorities:}
If the priorities of the organization shift over time due to increased requests based on a particular service, the managers will be able to reflect these changes through adjusting the weights of the evaluation metrics. 
Stakeholders and managers may have different opinions about the importance of a job type or activity, and a good performance evaluation system should allow the administrator and managers to investigate the impacts of applying different evaluation criteria.
\end{enumerate}

\section{\textcolor{firstround}{Deriving Performance Metrics}}\label{sec:data-flow}




\begin{figure*}[t]
  \centering
  \resizebox{\textwidth}{!}{\includegraphics{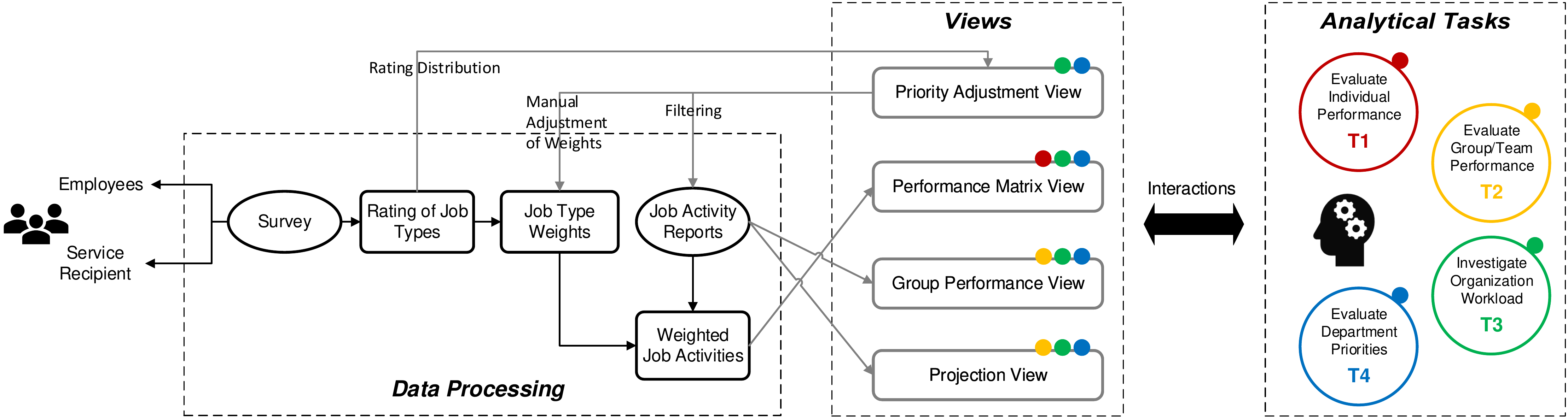}}
  \caption{Illustration of MetricsVis \textcolor{firstround}{system} diagram with three modules: data processing, views, and visual analytical task categories.}
  \label{fig:system-analysis-flow}
  \vspace{-3mm}
\end{figure*}

\begin{figure}[b]
  \centering
  \resizebox{0.95\columnwidth}{!}{\includegraphics{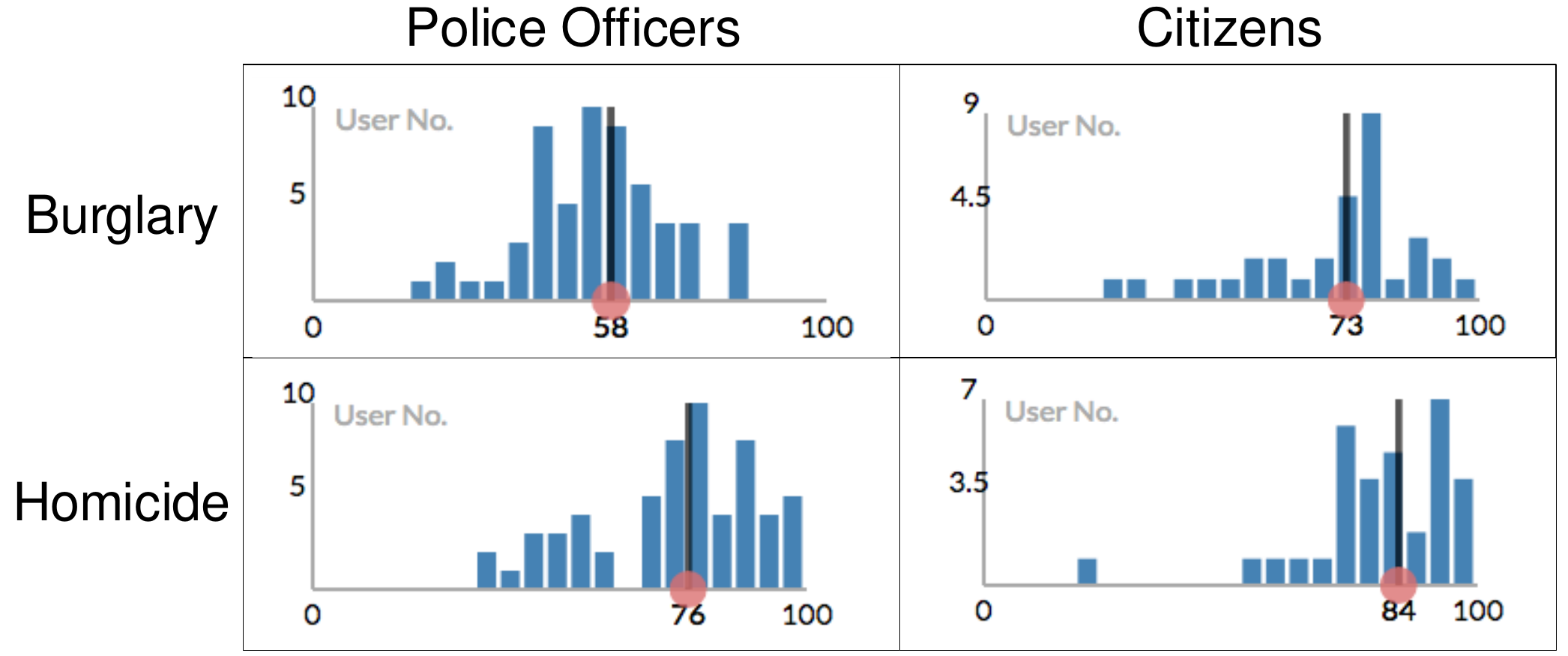}}
  \caption{The rating distribution of two sampled severe criminal offenses of burglary and homicide from police officers and citizens.
  In a histogram, the x-axis shows the rating scale from zero to one hundred, and the y-axis shows the count of each score. 
  The black lines denote the averages.}
  \label{fig:weight-rating}
  \vspace{-3mm}
\end{figure}


One of the key challenges in a successful performance appraisal system is quantifying the workload of employees and then deciding the contribution of specific jobs to the team or organizational objectives with appropriate scoring to reflect the priorities of a given organization.
We describe our method, which (a) transforms job activity reports to workload descriptors and (b) transforms subjective feedback from employees and communities to qualitative measurements of contributions (shown in Fig.~\ref{fig:system-analysis-flow} Data Processing). 
To demonstrate the performance data extraction stage, we utilize activity records of law enforcement agencies as an example and explain the transformation process in detail.
For that, we need to briefly describe the data source and several terms related to law enforcement agencies.

Law enforcement agencies typically use two related databases: computer-aided dispatch (CAD) and record management system (RMS).
The CAD tables contain calls for service event information such as call nature, address, time, patrol units dispatched, etc.
The calls usually fall into two categories: dispatched and self-initiated.
The self-initiated calls are usually started by officers on their patrolling duties, whereas dispatched calls are assigned to the officers.
RMS tables are concerned with criminal \textbf{incidents} that have been written into reports by officers, including parameters such as date, location, offense committed, etc.
We denote the calls ending with patrol service but not resulting in criminal incident report as \textit{Call for Service} \textbf{events}, which need to be considered as a separate category.
Dispatched activities and self-initiated activities should not be treated equally, since self-initiated activities are proactive behavior to prevent crimes and dispatched activities are responses to citizen requests;
the option of filtering activities by behavior types (self-initiated versus dispatched) is extremely useful for evaluating the performance of patrol officers.
\textcolor{calvin}{Both tables store data in a multi-dimensional format in which every entry contains a report with its associated metadata.}

Based on the taxonomy of major indicators of individual employee \textcolor{firstround}{task} performance~\cite{borman1993more,arvey1998performance,koopmans2011conceptual, cascio2011applied, Sanchez2011risk, campion2011competencies}, the top three common performance-related factors are \textbf{job completion, work quantity, and work quality}.
Rooted in these factors and based on feedback from our users, MetricsVis utilizes offense categories 
from law enforcement agencies as the diversity of job completion, the number of cases responded to by officers as work quantity, and the crowdsourced survey rating for the seriousness of offense categories as a practical substitute for work quality.
We conducted this crowdsourcing online survey with two participating groups: police officers and community citizens, each rating the severity and economic impact of each offense category on a Likert scale.
More details about the survey are available in our previous work~\cite{zhao2017metricsvis}. 
MetricsVis transformed these crowdsourced ratings to weights, which can be assigned either based on the average rating from the survey or on interactive adjustment from end-users.
Ultimately, the overall performance of an officer is calculated as the summation of these weighted offense categories.
In summary, the evaluation metrics are denominated as a set of hybrid evaluation metrics that contain (a) the quantitative measurement of employee achievements based on activity reports with respect to classified job types (data-driven) and (b) qualitative ratings from surveys or dynamic input from end-users (subjective input).

The ratings from citizens are much higher than those of officers for all offense categories. 
One possible interpretation is that since officers are exposed to a wide range of crimes on a daily basis, they have a less-biased viewpoint, whereas citizens usually only experience crimes from the position of a victim or witness.
Both officers and citizens weighted homicide as the top crime.
Fig.~\ref{fig:weight-rating} shows the rating comparison between police officers and citizens for \textit{Burglary} and \textit{Homicide}.
The rating from citizens has especially high spikes for both categories, but the rating from police officers is normally distributed. 

To compare the public's opinion with that of law enforcement towards different types of offenses, the chief from a partner law enforcement agency applied both of these weightings from officers and citizens.
He found that the ranking based on total performance score did not diverge significantly from their administrative goals. 
However, he noted the difference between citizens' concern towards some types of crimes and the officers' understanding of these crimes.
Although these differences did not affect the overall performance rating, they can be used in community meeting discussions to help align both groups' priorities.

\textcolor{firstround}{
Rating weights for other organizations can be obtained  directly from managers or supervisors.
Organizations can also survey employees and service recipients to obtain initial estimations on job category importance.
If it is easier for end users to rank the employees, initial weights can also be reverse-calculated using machine learning algorithms. 
However, weights obtained through such methods could be difficult to explain.
}
\textcolor{firstround}{As shown in Fig.~\ref{fig:system-analysis-flow}, the derived evaluation performance metrics and the job activity records are populated into designated views to show the performance of employees within and across multiple levels. 
}

\section{\textcolor{firstround}{MetricsVis System}}\label{sec:system}

MetricsVis is implemented as a web-based application that utilizes Redux~\cite{redux} to manage asynchronous calls between the client and server for data consistency, React~\cite{react} to support efficient updates of visualizations when data is modified, and D3~\cite{D3} to render the customized graphical interface. 
\textcolor{firstround}{MetricsVis contains four views: (1) a priority adjustment view displaying the domain-dependent evaluation metrics, (2) the performance matrix showing the details of individuals, (3) the group performance view showing the summarized results of groups as well as an individual's contribution to its group, and (4) the projection view supporting  similarity pattern analysis of team members.
In this section, each view is described based on its usage purpose and visual representations.
To demonstrate the visual representations in an example context, the views are rendered with datasets provided by a law enforcement agency.
}

\subsection{\textcolor{firstround}{Priority Adjustment View}}\label{sec:weight-table}

\textcolor{firstround}{The priority adjustment view} encodes the evaluation metrics that consider the diversity of evaluation factors in an organization.
Its main role is supporting dynamic selection of evaluation factors and adjustment of associated weights to match organizational priorities (T4).
As evaluation metrics in appraisal systems evolve due to rapid changes in service requests, we adopt a priority adjustment view to illustrate the contribution of each evaluation factor by associated weights.
After the dynamic modification of evaluation metrics (filtering of evaluation factors or tuning of weights), users can observe the impact on individual and group performance in all other views.

\begin{figure}[h]
  \centering
  \resizebox{0.9\columnwidth}{!}{\includegraphics{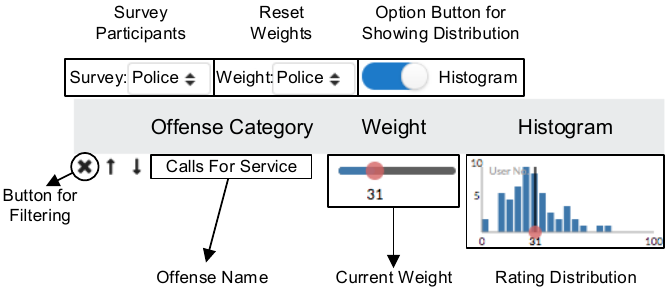}}
  \caption{A sample row in \textcolor{firstround}{priority adjustment view}: designed for law enforcement agencies.}
  \label{fig:sample_row_weight_table}
  \vspace{-0.5em}
\end{figure}

In the domain-dependent \textcolor{firstround}{priority adjustment view} (Fig.~\ref{fig:sample_row_weight_table}), each offense category appears as a row.
Each row has a slider bar to modify the current weight and a histogram to illustrate the rating distribution from either police officers or citizens.
In the histogram, the x-axis shows the rating weight scale from zero to one hundred, and the y-axis shows how many participants provided each rating score.
Placing the rating distribution beside the slider bar provides extra visual cues~\cite{willett2007scented} as to the severity of each offense category.
The rating distribution indicates the variation of opinions among survey participants.
The initial recommended weights are the average ratings from either police officers or citizens, with the exact value indicated by a red dot along the x-axis. 
Users can dynamically adjust the weight by dragging the red dot in a slider bar. 
If the priorities of the organization change, users can appropriately tune the weights of offense categories until the performance scores reflect the change in goals for the department.

\begin{figure}[t]
  \centering
  \resizebox{\columnwidth}{!}{\includegraphics{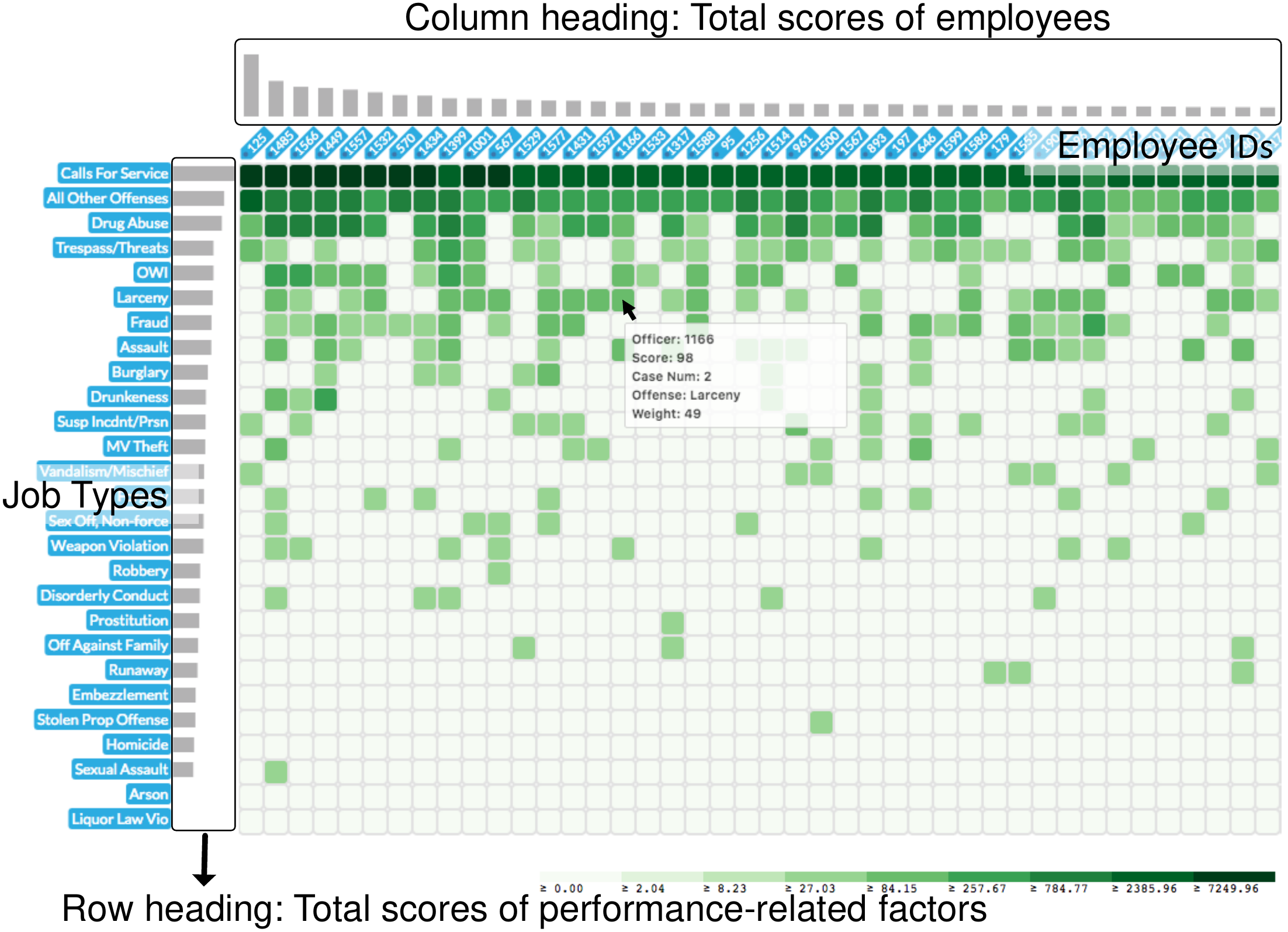}}
  \caption{The performance matrix shows the employees (columns) and job types (rows). The matrix is sorted based on the total score of employees and job types. Darker color encodes higher values.}
  \label{fig:performance_view}
  \vspace{-3mm}
\end{figure}

\subsection{\textcolor{firstround}{Performance Matrix View}}

\textcolor{firstround}{To efficiently evaluate and compare the performance of employees for the entire organization (T1, T3), our performance matrix (Fig.~\ref{fig:performance_view}) is designed to show the detailed job completion status of all employees in a holistic view.
We adopted a color-coded reorderable matrix for this purpose because the matrix (1) occupies a compact screen space to encode all employees, (2) provides a variation of a table to keep the familiarity, and (3) provides flexible sorting interactions.
In the matrix, employees and job types represent the columns and rows, respectively.}
The column heading located at the top with gray bars shows the total performance score of individuals. 
The row heading located at the left side shows the total score of performance-related factors. 
Each cell shows the performance score based on the hybrid evaluation metrics.
Each cell's color is defined by the sum of a job category accomplished by an employee.
When users mouse over a cell, a tooltip shows the precise score, number of completed jobs, and weight.
Clicking on an employee or job category re-sorts the table
with transitional animations.

\begin{figure}[t]
  \centering
  \resizebox{0.96\columnwidth}{!}{\includegraphics{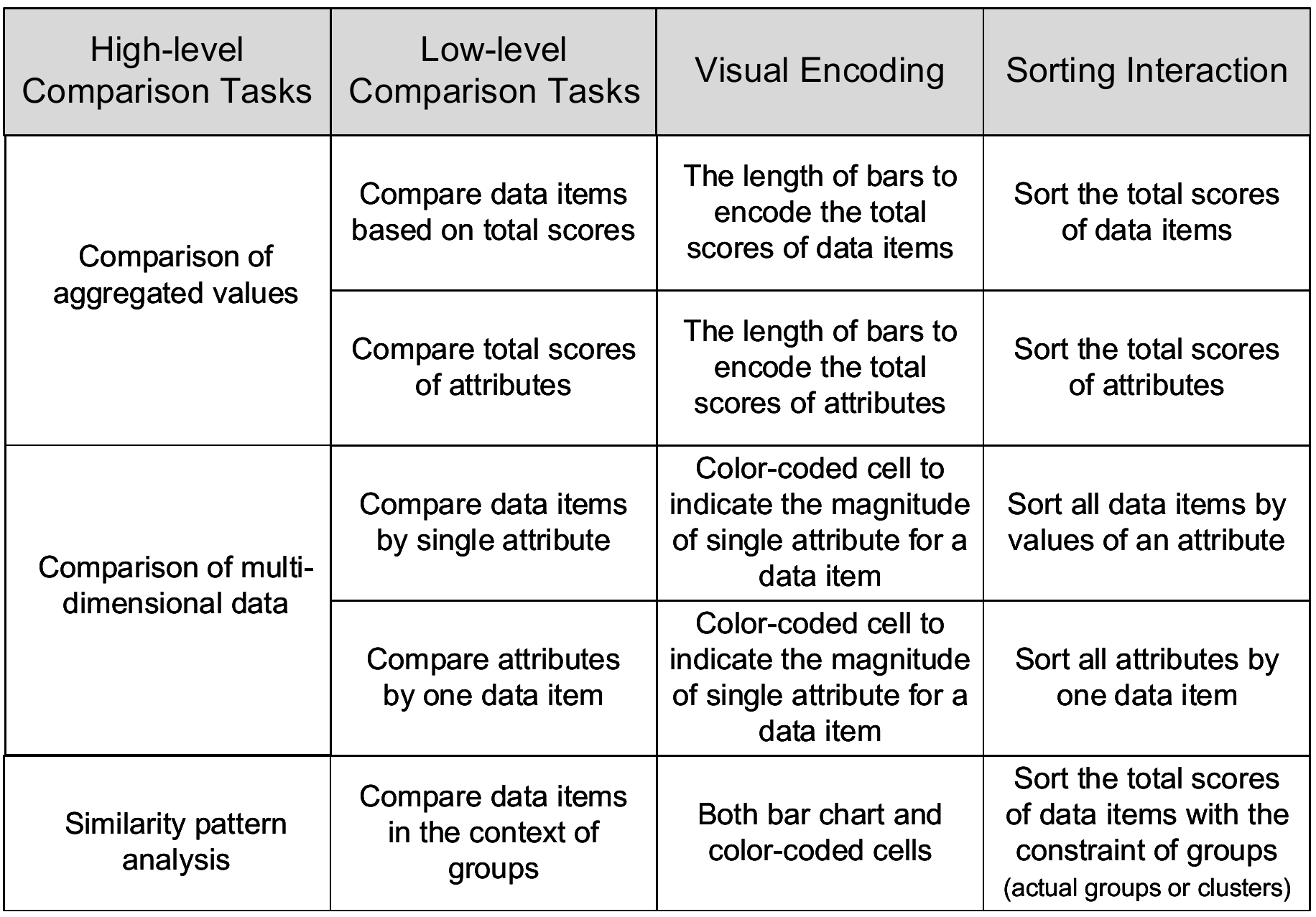}}
  \caption{The mapping of comparison tasks, visual encoding, and sorting interactions for the performance matrix view.}
  \label{fig:matrix_comparison}
  \vspace{-3mm}
\end{figure}

To satisfy the requirement of comparing multi-dimensional data at the individual level to the entire organization, the mapping of high- and low-level comparison tasks, visual encoding, and sorting interactions are listed in Fig.~\ref{fig:matrix_comparison}.
Two sorting interactions are provided in the performance matrix: (1) sort by total score of employees or job categories, and (2) sort by an individual employee or a particular job category.

We use event and incident records from a law enforcement agency as an example; the sorted results are shown in Fig.~\ref{fig:performance_view}.
Because some job types have low occurrence (e.g., arson, liquor law violation), the data in the performance matrix is relatively sparse.
In order to minimize the visual impact of this uneven data distribution and increase contrast within the matrix, we applied quantile mapping after a logarithmic transformation of the original scores and use nine sequential green colors recommended by ColorBrewer~\cite{colorbrewer}.
Our color mapping method is built on a data binning procedure that first normalizes the data using a power transformation and then applies equal interval binning on the transformed space~\cite{maciejewski2013boxcox}. 
We employed green colors because humans can perceive more shades of green than red or blue color tones~\cite{nicholls2001neuron}.

\begin{figure*}[tbh]
  \centering
    \resizebox{\textwidth}{!}{\includegraphics{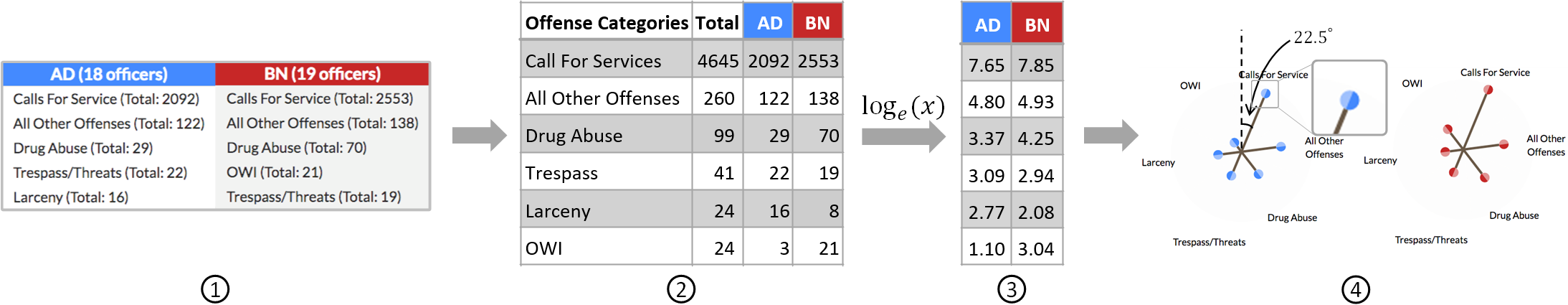}}
  \caption{The transformation steps from a table to dandelion glyphs. (1) Get the union of top five categories in both groups. (2) Order the categories by total in descending order. (3) Apply the logarithmic transformation to the total count. (4) Dandelion glyphs for two groups.
}
  \label{fig:table-dandelion}
  \vspace{-1.5mm}
\end{figure*}

As shown in Fig.~\ref{fig:performance_view}, the performance matrix is sorted by the total scores of officers in a descending order so that users can easily observe the officers with top performance scores.
Moreover, users can investigate the top performing officers who are dispatched, those who self-initiate the call response, or a combination of both (T1).
With offense categories sorted by a selected officer, users can observe the officers' relative workload across different offense categories and where they focused their self-initiated work. 
This helps commanders understand the strengths and weaknesses of an officer.
The performance matrix includes all members across the organization, which provides comparisons in an organizational context. 
Users can observe how an officer ranks in the organization. 
Selection interactions are supported to simplify officer comparison; for instance, users can select any officers that they are interested in and then those officers will be aligned on the left side of the matrix. 
With selection operations, commanders can evaluate and compare the officers in their teams and explore the different types of incidents responded to by individuals, their team, and the organization.

Sorting by total score of offense categories demonstrates the overall workload needed to be addressed by an agency (T3).
Comparing the total score of offense categories in two time frames, such as between consecutive months, can indicate changes in the prevalence of certain crimes. 
Ranking officers by a given offense category can directly reveal the most experienced officers for dealing with such incidents.
If the police department wants to target a specific crime category, the commanders can determine the officers most suited for the task.

\begin{figure}[t]
  \centering
  \resizebox{\columnwidth}{!}{\includegraphics{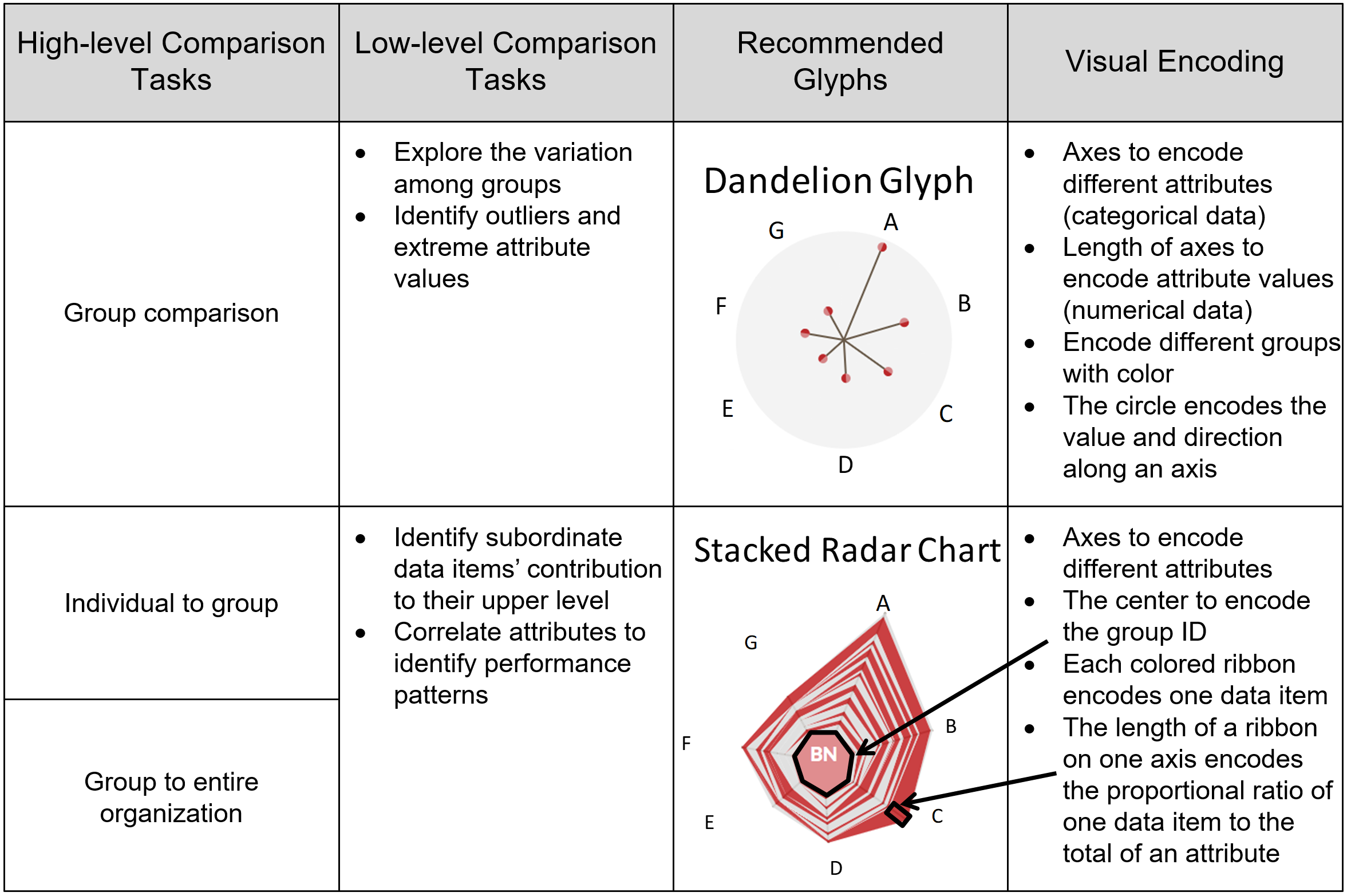}}
  \caption{The mapping of comparison tasks, glyphs, and visual encoding for the group performance view.}
  \label{fig:group_comparison}
  \vspace{-4mm}
\end{figure}

\subsection{\textcolor{firstround}{Group Performance View}}\label{subsec:group_performance}


Most organizations have employees working in teams; as a result, for effective performance evaluation, it is essential to understand the performance among these groups.
To support comparisons among groups (T2), our system provides two grouping methods: (1) group by team assignments and (2) group by a clustering algorithm.
We implemented three visual representations in our overall group performance view to support this comparison and analysis of team performance: (1) a table list, (2) dandelion glyphs, and (3) stacked radar charts.
The group performance view provides an overview of the aggregated multi-dimensional performance data items for groups.
For high-level comparison tasks (Fig.~\ref{fig:group_comparison}), the group performance view demonstrates (1) performance evaluation and comparison at the group level (\textcolor{firstround}{within the same level}), and (2) each individual data item's performance contribution to its group and performance contribution of a group to the entire organization (\textcolor{firstround}{across two levels}).
For low-level comparison tasks, the customized dandelion glyphs provide an efficient simultaneous comparison for a set of data attributes, and identification of outliers and correlation among attributes.
The combination of the dandelion glyph (Section~\ref{subsubsec:dandelion_glyph}) and stacked radar chart (Section~\ref{subsubsec:stacked_radar_chart}) enables retention of inherent hierarchical relationships among employees and supports high- and low-level comparison tasks.
In addition, the dandelion glyph displays an overview of a group, with the details of individual employees expanded on-demand in the stacked radar chart.

\subsubsection{Table}\label{sec:group_table}
The table at the top of the group performance view shows the overall performance of each 
group (\textcolor{firstround}{Fig.~\ref{fig:table-dandelion}(1) group by assigned shifts}). 
With a summarization of jobs accomplished by employees, the table lists the ranking of job categories based on their counts, making it intuitive for users to examine the workload of each team.
For instance, patrol officers that work in law enforcement agencies need to constantly monitor designated areas to ensure the safety of the community and are usually assigned by shifts and districts.
The \textit{A shift} and \textit{B shift} split the days of week (alternating days). 
Each day is broken down into a day shift and a night shift. 
Fig.~\ref{fig:table-dandelion}(1) shows the top five offense categories for \textit{A Day shift}\raisebox{0pt}[0pt][0pt]{\raisebox{-1.4ex}{\includegraphics[height=3.4ex]{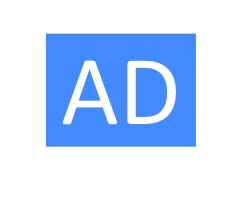}}}and \textit{B Night shift}\raisebox{0pt}[0pt][0pt]{\raisebox{-1.4ex}{\includegraphics[height=3.4ex]{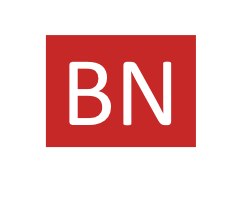}}}. 
For both teams, officers spent the most time on \textit{Calls for Service} events that did not generate criminal case reports or incidents belonging to the \textit{All Other Offenses} category. \textit{Calls for Service} events are not considered a high priority, but generate a large portion of the workload. 
Our law enforcement partner agency found that this view provided the insight that they needed to break down the \textit{All Other Offenses} category and examine which offenses in this category should receive further examination. 
With the ordering of offense categories for each group, users can easily find out which tasks utilize the most resources from each team and shift. 
However, it is not as easy to compare the different groups with only the table listing. 
Thus, we created a dandelion glyph to enable convenient comparison among such groups.

\begin{figure*}[tbh]
  \centering
  \resizebox{\textwidth}{!}{\includegraphics{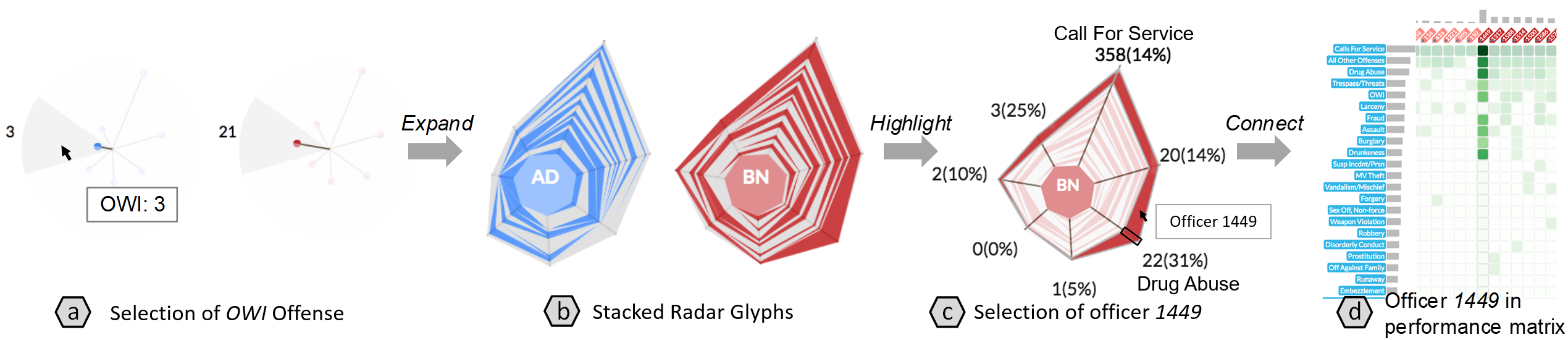}}
  \caption{The two radial layout visual representations in the group performance view: dandelion glyphs and stacked radar glyphs.
  The glyphs show the list of criminal incidents that responded by \textit{A Day shift} and \textit{B Night shift}. (a) Highlight of \textit{OWI} incidents in dandelion glyphs. (b) Stacked radar glyphs show the contribution of each member. (c) Selection of Officer \textit{1449} in \textit{B Night shift}. (d) Highlight of Officer \textit{1449} in performance matrix.}
  \label{fig:group_performance_view}
  \vspace{-1.5mm}
\end{figure*}

\begin{figure}[t]
  \centering
  \resizebox{\columnwidth}{!}{\includegraphics{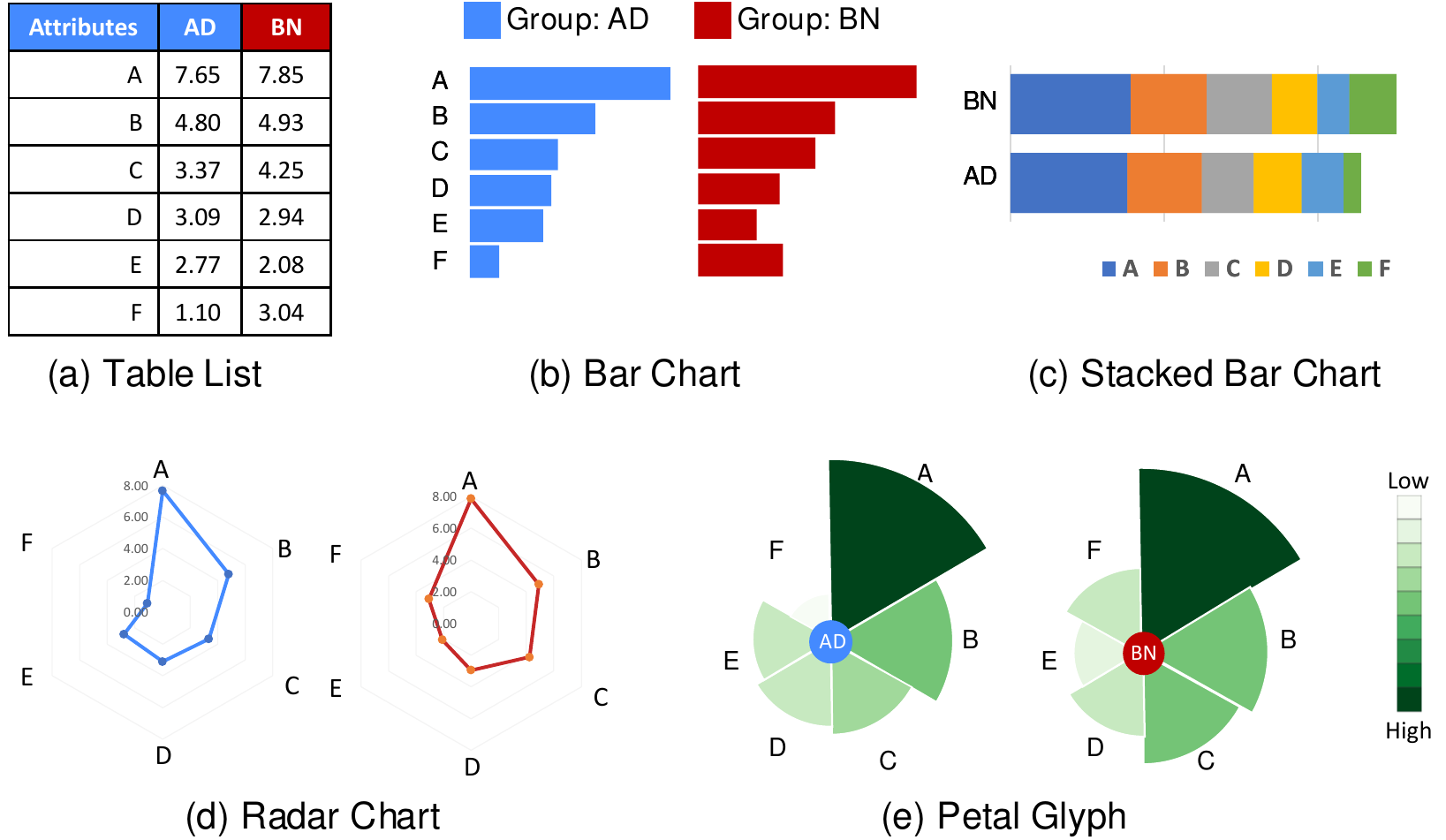}}
  \caption{The five examples of small multiples glyph to represent the multi-dimensional data attributes of two groups.}
  \label{fig:examples_visual_groups}
  \vspace{-3mm}
\end{figure}

\subsubsection{Dandelion Glyph}\label{subsubsec:dandelion_glyph}
Small multiple glyphs are expressive and use screen space effectively to illustrate large data~\cite{fuchs2013evaluation}.
Thus, we incorporated characteristics of various small multiple glyph designs (Fig.~\ref{fig:examples_visual_groups}) into the design of the dandelion glyph.
Inspired by previous research indicating that star plots with radial layout outperform tabular displays for comparing attribute values~\cite{keck2017glyph}, we also adopted the radial layout into our dandelion glyph. 
In our dandelion glyph, the axes encode different attributes (categorical data) and the length of the axes encode the attribute values (numerical data). 
We compare our dandelion glyph with the graphs shown in Fig.~\ref{fig:examples_visual_groups} and Table~\ref{tab:dandelion_glyph_vs_others}. 
Dandelion glyphs have high data-to-ink ratio and are intuitive, visualizing the differences among groups effectively.



The transformation process from tabular display to dandelion glyphs is shown in Fig.~\ref{fig:table-dandelion}.
In step 2, we took the union of offense categories across all groups, and then ordered the offense categories based on the total count.
Finally, the order of axes was determined based on the total count of each attribute.
A logarithmic transformation of the total count was applied to the dandelion glyph, since the values of each category axes vary extensively in our dataset.
\textcolor{luke}{However, datasets with minimal variance (e.g., agencies in which group performance categories contain similar values) might only require linear transformations.}
The transformation enables users to perceive the contribution of each job category.
Notably, the dandelion glyph is a simplified version of star coordinates. 
Munzner~\cite{munzner2014visualization} discussed the suitable scenarios of applying radial layout and importantly mentioned the inappropriate representation effect that symmetric axes have on the same value.  
To eliminate symmetric impressions in our dandelion glyphs, we rotated the glyph by a small amount ($\frac{1}{8}\pi$). 

\textcolor{firstround}{Although the dandelion glyph is most suitable for displaying up to 10 to 12 attributes, users can interactively adjust the number of top categories in each group.}
Users also can interactively explore the total count among groups with selection interaction. 
Comparing two groups' performance in Fig.~\ref{fig:table-dandelion}(4), the significant difference of \textit{OWI} incidents is readily apparent. 
To further confirm the exact numerical difference, users can select the \textit{OWI} axis, and the corresponding axes in all dandelion glyphs are highlighted with precise values (Fig.~\ref{fig:group_performance_view}(a)). 
The dandelion glyphs represent an overview to avoid initial visual clutter and can be expanded to stacked radar charts to show moderate details of individuals on-demand.

\subsubsection{Stacked Radar Chart}\label{subsubsec:stacked_radar_chart}
The stacked radar chart is customized to illustrate contributions of subordinate individuals to their upper levels/groups, and it holds the same contour as dandelion glyph to keep familiarity and consistency.
It can be applied to show connections between two levels and preserve the information of aggregated upper groups and show moderate details of subsequent levels in a compact space.
An example of a stacked radar chart can be found in Fig.~\ref{fig:group_performance_view}(b).
For a group member, the values of axes are shown as colored ribbons in the radial layout. 
As shown in Fig.~\ref{fig:group_performance_view}(c), the selected officer \textit{1449} dealt with 22 \textit{Drug Abuse} incidents, which is around $31.42 \%$ ($\frac{22}{70}$) of the entire group. 
The proportion of pixels along one axis is calculated based on the ratio between the value of a member to the group total. 
Using the link to the performance matrix, we can observe that, unsurprisingly, officer~\textit{1449} had the top performance score in his or her group (Fig.~\ref{fig:group_performance_view}(d)).

The stacked radar chart allows users to inspect the contribution ratio of each member of a group.
\textcolor{firstround}{
Diehl et al.~\cite{diehl2010radial} found that using a radial layout to encode data attributes by sectors outperforms Cartesian coordinates (i.e., matrix) when focusing on one dimension.
This observation from Diehl et al. was made based on an evenly distributed radial grid layout with a single grid highlighted.
In our scenario, colored ribbons are adopted to show the variations across multiple attributes simultaneously.
We chose to use this method because it allows users to not only identify which members contribute significantly to a group, but to compare performance pattern with those of other members as well.
}
However, while the stacked radar chart effectively demonstrates individual contributions within a group context, users should use it with caution.
The length of axes cannot be compared directly since a logarithmic transformation (a non-linear monotonic function) is applied in the dandelion glyph generation process, yet values within an axis are linearly mapped.
As discussed in the previous section, the transformation is necessary due to the skewed nature of the original input dataset. 
Our approach is a tradeoff between encoding the actual value and providing appropriate visual perception. 
In conclusion, we believe the advantage of using stacked radar charts outweighs the side effects caused by the transformation.
To compensate for the uneven spatial distribution of the radial layout, we add a null inner circle (Fig.~\ref{fig:group_performance_view}(c)) to reduce the bias introduced in the connection between axes.

\begin{table}[t]
\small
\centering
\caption{Comparison of dandelion glyph versus other glyphs in small multiple settings.}
\smallskip
\begin{tabular}{lll}
\hline
\rowcolor[HTML]{D3D3D3} 
Visualization & Advantages & \multicolumn{1}{l}{\begin{tabular}[c]{@{}l@{}}Disadvanrages for \\ Increased Data Size \end{tabular}}  \\ \hline
\multicolumn{1}{l}{Dandelion Glyph} & \multicolumn{1}{l}{\begin{tabular}[c]{@{}l@{}}$\bullet$ Radial layout$^1$\\ $\bullet$ High data-ink ratio \end{tabular}} & \multicolumn{1}{l}{\begin{tabular}[c]{@{}l@{}}$\bullet$ Scalability problem$^3$ \end{tabular}} \\ \hline
\multicolumn{1}{l}{Table List} & \multicolumn{1}{l}{\begin{tabular}[c]{@{}l@{}}$\bullet$ Precise values\\ $\bullet$ Commonly understood\end{tabular}} & \multicolumn{1}{l}{\begin{tabular}[c]{@{}l@{}} $\bullet$ Less efficient in \\ \SmallIndent comparison tasks\\ $\bullet$ Large pixel size for \\ \SmallIndent single data item\end{tabular}} \\ \hline
\multicolumn{1}{l}{Bar Chart} & {\begin{tabular}[c]{@{}l@{}} $\bullet$ Rectangular layout$^2$\end{tabular}} & \multicolumn{1}{l}{\begin{tabular}[c]{@{}l@{}} $\bullet$ Complex in comparison tasks \\ \SmallIndent than radial layout: harder to \\ \SmallIndent locate identical attribute\end{tabular}} \\ \hline
\multicolumn{1}{l}{Stacked Bar Chart} & \multicolumn{1}{l}{\begin{tabular}[c]{@{}l@{}} $\bullet$ Easy to compare the \\ \SmallIndent sum of all attributes\end{tabular}} & 
\multicolumn{1}{l}{\begin{tabular}[c]{@{}l@{}} $\bullet$ Hard to compare the bars \\ \SmallIndent in the middle\end{tabular}} \\ \hline
\multicolumn{1}{l}{Radar Chart} & \multicolumn{1}{l}{$\bullet$ Radial layout$^1$} & \multicolumn{1}{l}{\begin{tabular}[c]{@{}l@{}}$\bullet$ Scalability problem$^3$\\ $\bullet$ The connections at the end \\ \SmallIndent of axes are unnecessary\\ $\bullet$ Unequal importance \\ \SmallIndent among attributes\end{tabular}} \\ \hline
\multicolumn{1}{l}{Petal Glyph} & \multicolumn{1}{l}{\begin{tabular}[c]{@{}l@{}}$\bullet$ Radial layout$^1$\\ $\bullet$ Double encoding (length \\ \SmallIndent and color) for values\end{tabular}} & \multicolumn{1}{l}{\begin{tabular}[c]{@{}l@{}} $\bullet$ Scalability problem$^3$\\ $\bullet$ More pixels on the screen \\ \SmallIndent are used for each attribute\end{tabular}} \\ \hline
\end{tabular}
\label{tab:dandelion_glyph_vs_others}
\footnotesize{$^1$ Efficient in comparison tasks for large data~\cite{keck2017glyph}, $^2$ Simple layout to indicate the data variance, $^3$ Only appropriate to show a dozen data attributes or less~\cite{munzner2014visualization}}\\
\vspace{-3mm}
\end{table}

\textcolor{firstround}{
Compared with matrix and tabular visualizations, the stacked radar chart is less precise regarding showing exact values.
The alternating neighboring colors are used to separate individuals in a compact screen space; therefore, only a limited number of items can be shown.
Filtering interactions (showing only a few of the members) and keyboard selection of a single data item mitigate the scalability issues of the stacked radar chart.
In our informal interview with domain experts, they confirmed the advantages of using stacked radar chart as following: easy and quick identification of high-contributing individuals and extreme attribute values.
}

\subsection{\textcolor{firstround}{Projection View}}

The projection view contains a scatterplot showing the projected distance among data items. 
In this view (Fig.~\ref{fig:teaser}(b)), each data item is shown as a solid dot with an identifiable label, and its color encodes the group information.
For instance, the officers close to each other in the projection view have handled similar types of offenses, and their performance is highly correlated.
During shift planning, team commanders can build a new team of employees with similar experiences addressing specific types of crime.

To assist with designing resource allocation strategies that balance workload and the skill set of a group, we applied a K-means clustering method~\cite{hartigan1979algorithm}.
\textcolor{firstround}{The scatterplot displays the results of a manifolds dimensionality reduction algorithm t-SNE~\cite{maaten2008visualizing}, which can reduce the multi-dimensional data into  a lower number of dimensions to reveal the relationship among data items.
The clustering results are marked in the scatterplot through group colors.}
Users may adjust the number of clusters to get rid of outliers, since the K-means algorithm is sensitive to noise.
For K-means, the input data attributes are the number of cases in offense categories.
A normalization of input data (maps the original range of one attribute to the range 0 to 1) is applied to guarantee each data attribute contributes properly to the final clustering results.

\section{Evaluation}\label{sec:evaluation}

To demonstrate how our partner agency utilized MetricsVis in exploring event and incident records as well as evaluating patrol officers' effectiveness, we describe two example use cases.

\subsection{Use Case 1}\label{subsec:case1}

The chief of a law enforcement agency needs to build provisional specialized anti-drug teams.
Before forming the new teams, he wants to know the historical workforce performance of handling drug abuse incidents.
He is interested in exploring five months (July 1st to Dec 1st, 2017) of incident records (Fig.~\ref{fig:teaser}(1)).
He selects all officers and filters out the dispatched cases and call for service events (Fig.~\ref{fig:teaser}(2)),
since the majority of drug abuse incidents are self-initiated and result in a criminal report.

He first examines the ranking of officers' total scores in the performance matrix view.
He observes that the top 3 officers responded to 88, 74, and 67 total cases, respectively.
Next, he examines the most prevalent crimes through sorting by offense categories. 
He finds that drug abuse is the second most frequent offense category (Fig.~\ref{fig:teaser}(3)) with an initial weighting of 64 (average rating from police officers). 
In examining the precise numbers, he notices that the top 3 officers handled 36, 33, and 15 drug abuse incidents, which takes up $40.90\%$, $44.59\%$, and $22.39\%$ of their self-initiated workload.
To explore drug abuse cases more closely, he directly sorts the officers by drug abuse offense category and discovers that 52 officers were involved in a total 383 cases (ranging from 1 to 36 by individual officer).
With 3 officers handling over 20\% of the cases,
when creating an anti-drug team, these officers and officers with similar performance across all cases are good potential candidates.

Since offense categories are not independent and drug abuse is highly correlated with $80\%$ of crimes, he explores the activity patterns of the 52 officers more closely using the automatic grouping generated by our clustering algorithms and the visualization results in the group performance view, projection view, and performance matrix view.
After a few trials, he finds that K-means clustering with six clusters (Fig.~\ref{fig:teaser}(4)) provides a good grouping of the results to understand the performance pattern among officers. 
The majority of the 52 officers are scattered into four clusters, and officers in three clusters responded to majority of the total number of drug abuse incidents: the blue cluster \raisebox{0pt}[0pt][0pt]{\raisebox{-1.4ex}{\includegraphics[height=3.4ex]{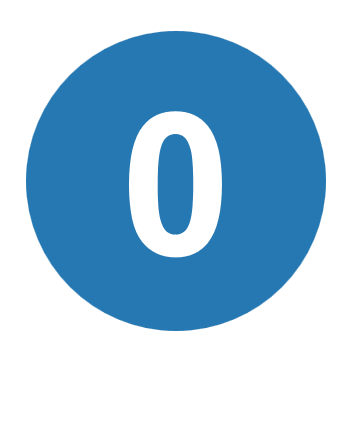}}} of 11 officers handled 133 drug abuse incidents, the red cluster \raisebox{0pt}[0pt][0pt]{\raisebox{-1.4ex}{\includegraphics[height=3.4ex]{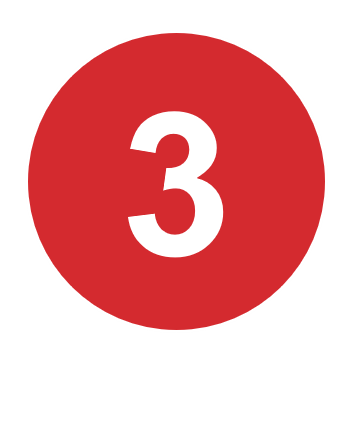}}} of 7 officers handled 97 drug abuse incidents, and the brown cluster \raisebox{0pt}[0pt][0pt]{\raisebox{-1.4ex}{\includegraphics[height=3.4ex]{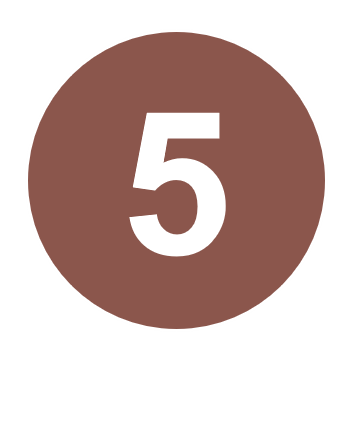}}} of 12 officers handled 84 drug abuse incidents (Fig.~\ref{fig:teaser}(5)). 
The combination of these three clusters of 43 officers responded to 81.98~\% of 
drug abuse incidents. 
\textcolor{firstround}{He also inspects the clustering results in the projection view to observe the similarity pattern among clusters, where he finds that green \raisebox{0pt}[0pt][0pt]{\raisebox{-1.4ex}{\includegraphics[height=3.4ex]{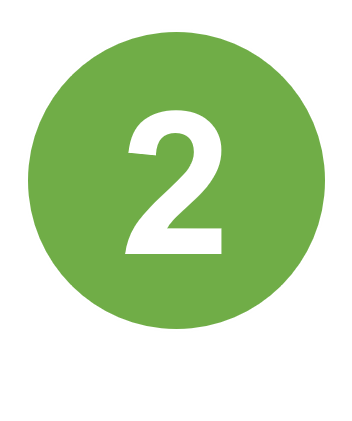}}} and red \raisebox{0pt}[0pt][0pt]{\raisebox{-1.4ex}{\includegraphics[height=3.4ex]{red_cluster}}} clusters are farther apart in projection space than blue \raisebox{0pt}[0pt][0pt]{\raisebox{-1.4ex}{\includegraphics[height=3.4ex]{blue_cluster}}} and brown \raisebox{0pt}[0pt][0pt]{\raisebox{-1.4ex}{\includegraphics[height=3.4ex]{brown_cluster}}} clusters, which can also be observed in the group performance view (Fig.~\ref{fig:teaser}(6)).}
He digs into the details among these 3 clusters by first examining the dandelion glyphs
(Fig.~\ref{fig:teaser}(7)). 
Besides large numbers of overlapping cases (e.g. trespass/threats, operating [a vehicle] while intoxicated (OWI)),  he finds that officers in the blue cluster \raisebox{0pt}[0pt][0pt]{\raisebox{-1.4ex}{\includegraphics[height=3.4ex]{blue_cluster}}} also dealt with many larceny cases, officers in the red cluster \raisebox{0pt}[0pt][0pt]{\raisebox{-1.4ex}{\includegraphics[height=3.4ex]{red_cluster}}} dealt with many burglary cases, and officers in the brown cluster \raisebox{0pt}[0pt][0pt]{\raisebox{-1.4ex}{\includegraphics[height=3.4ex]{brown_cluster}}} handled many assault cases. 
The stacked radar chart (Fig.~\ref{fig:teaser}(8)) shows the patterns among the three groups and distinguishable officers in each group.
By further examining the officers in the performance matrix (Fig.~\ref{fig:teaser}(9)), the chief identifies the officers that are most experienced with combinations of different offenses with drug abuse.
\textcolor{calvin}{``This tool provides [commanders] with objective data to assist in resource deployment decision making rather than solely relying on subjective, `best guess', practices that are the norm in law enforcement,'' commented the chief.}

\subsection{Use Case 2}\label{subsec:case2}

The department currently evaluates each officer by their supervisors' scores, which contain subjective metrics that are time-consuming and possibly biased.
The chief wants to know if he can utilize data-driven officer metrics in combination with MetricsVis to more effectively and efficiently evaluate performance.
He applies the average weighting initially provided by police officers to each incident type.
He uses the same time frame as Use Case 1, as well as all call events and crime incidents for both dispatched and self-initiated activities. 
He now compares his view and his command staff's view of the top performing officers versus the results shown in our system.
Interestingly, the top-ranked officer in the performance matrix view does not match their internal evaluation results. 
Interactively exploring factors and ideas about what they consider characteristics of the best officers, he decides to consider only criminal incidents that exclude the call events. 
He finds some officers that match his understanding of good performance get better rankings in the performance matrix under this system. 
Exploring deeper, he proceeds to filter out dispatched incidents, because he thinks self-initiated incidents are a key component of a top officer.
He finally finds that a ranking using only self-initiated incidents matches his command team's understanding of top individual officer performance.

\begin{figure}[ht]
  \centering
  \resizebox{\columnwidth}{!}{\includegraphics{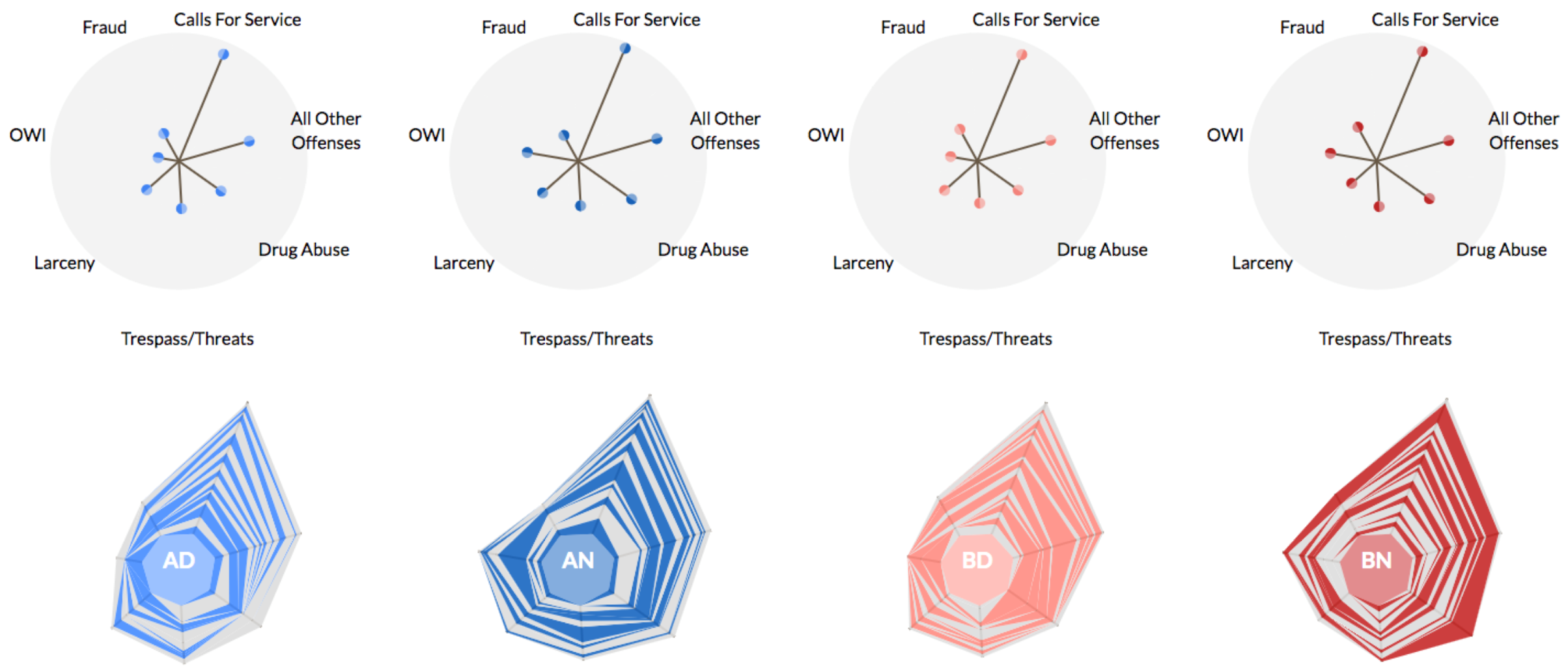}}
  \caption{Day (\textit{AD}, \textit{BD}) and night (\textit{AN}, \textit{BN}) shifts have significant differences in drug and OWI incidents for self-initiated incidents.}
  \label{fig:case2-1}
  \vspace{-3mm}
\end{figure}



With the confirmation of the effectiveness of the collected evaluation metrics, the chief is interested in investigating the difference between shifts and districts.
He continues with shifts grouped using self-initiated incidents. 
(As mentioned in Section~\ref{sec:group_table}, \textit{A shift} and \textit{B shift} are alternating by days, and each day is broken into a day shift and a night shift. Some patrol officers are not assigned to a specific shift.)
It is not surprising that the day shifts exhibit a similar pattern and the night shifts show another trend (Fig.~\ref{fig:case2-1}). 
Based on the dandelion glyphs, he notices that the significant difference between day shift and night shift is the number of drug abuse and  OWI cases. 
He also wants to compare the dispatched incidents between shifts, and expects the four shifts to have very similar patterns and the workload to be evenly distributed across all shifts for dispatched incidents. 
He also wonders about the workload across different districts. 
Even for dispatched incidents, the difference is noticeable. 
Therefore, these differences can be used to guide effective policing on each shift and district and also must be factored into an officer's performance evaluation, since an officer should not be scored poorly because they are assigned to a low crime time period and area.
\textcolor{calvin}{A lieutenant from highway patrol recognizes this: ``MetricsVis would enable [commanders] to look at the total impact of officers and teams and not just sums of cases/incidents. This enables them to assess team and organization level performance in achieving their goals.''}

\section{Domain Expert Feedback}\label{sec:feedback}

We deployed the system to a local police chief, shift commanders, and a crime analyst. 
The local chief stated that MetricsVis is a valuable visual analytics tool that supports a broad view of the entire organization and provides the possibility to break stereotypes and overcome bias in understanding organizational performance. 
MetricsVis has also revealed new insights into staff workload and which quantitative metrics (e.g. self-initiated incidents) relate to supervisor's subjective evaluation of top officers.
Moreover, the chief noticed the necessity of deconstructing the All Other Offenses category, which contains around 50~\% of criminal incidents.
Drilling down on this generic offense category can improve the comprehensiveness of evaluation metrics in aligning with organizational objectives.

The command staff have now used the tool during their last four quarterly performance reviews and have provided very positive feedback.
They expressed that the tool enables them to ground their evaluations, and quickly and effectively explore understandable quantitative metrics.
It also indicates role models and activity types for officers to use as guidelines for improving their performance.
Another noted valuable aspect of MetricsVis is its ability to convey the most effective and experienced officers for handling certain incident types.
This information is helpful in preparing shifts and training sessions.

A crime analyst who was engaged in the development process of MetricsVis provided valuable interpretation of the data (e.g. night shifts often deal with more self-initiated incidents even though there are fewer calls after midnight, since officers during the day are largely occupied with dispatched cases;
day and night shifts usually have very different working patterns), as well as helped validate datasets and define questions of interest.
He has identified additional factors that contribute to organizational performance for inclusion for future improvement of MetricsVis (e.g. days worked, arrests, traffic stops).




\begin{figure}[t]
  \centering
  \resizebox{\columnwidth}{!}{\includegraphics{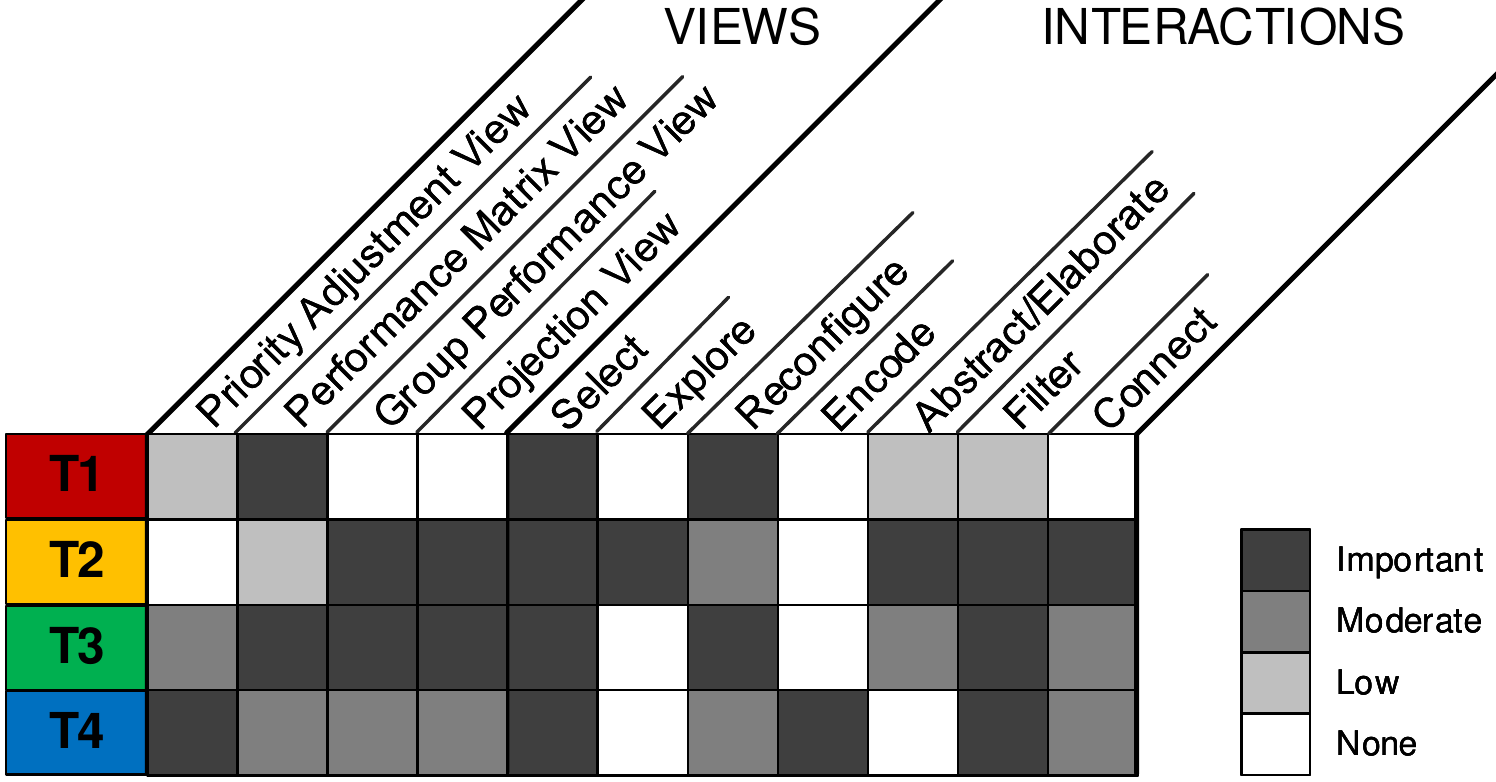}}
  \caption{The relationship between analytical tasks (rows) and MetricsVis views and low-level interaction categories~\cite{yi2007toward} (columns). 
  Cell shading quantifies how a particular view or an interaction contributes to the analysis process of a task. 
 }
  \label{fig:task_view_interaction}
  \vspace{-3mm}
\end{figure}

\section{Discussion}\label{sec:discussion}
\paragraph{Tasks, Views, and Interaction Mapping} 
To accomplish each task, a number of views and several interaction categories (proposed by Yi et al.~\cite{yi2007toward}) are required.
Fig.~\ref{fig:task_view_interaction} outlines the role of the views and interactions needed for each task, \textcolor{firstround}{and the shaded cells were colored based on the frequency of using views and interaction categories to accomplish tasks during the interactive sessions with domain experts (police and commanders)}.
To efficiently evaluate individual employee performance (T1), the performance matrix view is frequently used to explore the details of all employees in a holistic view.
\textcolor{luke}{Users can highlight (select) a subset of employees to compare and rank by performance with sorting and reconfiguring interactions.}
To support comparisons among groups (T2), the group performance view shows the aggregated results of groups (group level comparison) as well as the contribution of group members (across individual and group level comparison).
Abstract/elaborate interaction categories are frequently used to show the overview among groups first and on-demand details of individuals within one group in the group performance view.
Select, explore, filter, and connect are the basic interaction categories for linking employees to their groups and identifying prominent patterns (e.g.. anomalies with low/high performance).
Evaluating organizational workload (T3) and priorities (T4) are more comprehensive tasks that require exploration with all views.
Analyzing the workload across the entire organization (T3) involves the summation of all completed jobs during a certain period.
The performance matrix aggregates all jobs completed by selected employees, showing the productivity outcome of the entire organization.
The options to select, filter, and reconfigure interaction categories provide the flexibility to investigate the overall performance over different time frames, locations, and alternative team assignment.
To verify the alignment of evaluation metrics vs. department priorities (T4), the priority adjustment view is heavily used for the filtering of job types (filter) and tuning of weights, and then the corresponding changes are reflected in all other views (encode).

\paragraph{Evaluation Metrics} 
We started by trying to understand the general characteristics of employees, teams, and shifts with different workloads in various organizations while consulting the literature.
Our collaboration with domain experts from law enforcement agencies enabled us to better understand the importance of refining the evaluation metrics.
We considered using the number of responded cases per officer to represent the quantitative measurement of productivity.
However, the effort required to resolve each case is different.
After consultation with police supervisors, we adopted the idea of substituting the effort of handling a case with the severity of the crime.
The severity somewhat reflects the relative importance of responding to a case.
Based on the initial weights determined through surveys, domain experts can dynamically investigate the overall performance, which is derived using additive weighting.
The goal of MetricsVis is not micromanagement (deciding who is the best officer), but a systematic approach to investigating the effectiveness of an organization at  and across multiple levels.
The optimization of evaluation metrics is an ongoing area of research, but with the assistance of MetricsVis, domain experts can investigate different sets of evaluation metrics to identify the best match with their organizational objectives.

\paragraph{Visual Designs} 
For visualization design, we deduced that a tabular visualization summarizing all employees provided better utility than graphs of individual statistics.
Besides the individual performance, we noticed the importance of providing overview on the aggregated group results.
After examination of different designs in small multiple settings, we adopted a dandelion glyph, which is a variation of star coordinates.
We added the stacked radar chart to bridge the gap between dandelion glyph (group) and performance matrix (individual).
The stacked radar chart shows all members within a group as well as their performance-related factors in limited screen space. 
\textcolor{firstround}{Compared with treemaps~\cite{shneiderman1998treemaps} and node-linked graphs that usually focus on displaying the hierarchical relationships among data items, the stacked radar chart allows simultaneous comparison for multiple attributes using continuous shape instead of separated ones.}
The radial layout can show only a limited number of visually differentiable categories; however, the number of common job types across different teams is limited, \textcolor{firstround}{and filtering interactions and selection by keyboard can improve the usability}. 

\paragraph{Generalization}
We believe that the four visual analytical tasks categories identified in this paper are applicable to other team- or shift-based organizations that use automatic systems to record employee activities, such as delivery drivers, nurses, and emergency medical services.
\textcolor{luke}{In addition, MetricsVis, although implemented for public safety agencies, was designed with individual and group performance evaluation in mind and, therefore, we expect that the system can be extended to similar type organizations.}
\textcolor{firstround}{
}

\paragraph{Limitations} 
There are several limitations in our current system. 
For instance, officers who work fewer shifts cannot be directly compared with officers working full shifts. 
Also, the number of hours officers work each shift is not currently logged.
The time required to respond to each type of incident needs to be incorporated as a weighting factor when computing metrics of performance.   
Currently, our system is designed for organizations with only a few hundred employees and dozens of job categories.
Scalability of the system for larger organizations may be an issue as the number of  dimensions for similarity pattern analysis increases;
additional hierarchical modeling and filtering may be a solution for scaling to higher dimensions.

\section{Conclusion and Future Work}\label{sec:conclusion}
We presented MetricsVis, an interactive visual analytics system for organizational performance evaluation.
Our system contains four visual components to support interactive visual analysis of organizational performance with a set of hybrid evaluation metrics, integrating subjective ratings and quantifiable outcomes of job activities at multiple grouping granularities.
The usability of MetricsVis was demonstrated with two use cases that leverage the designed features and their use for real-world problems: new group staffing and actual group assignments to shifts and districts.

To optimize and improve the evaluation metrics, we plan to incorporate more activity records (e.g. number of arrests, traffic stops).
Another possible improvement is to include the time associated with job types as another contributing factor in the final performance outcome, since the time to complete a particular problem is of interest regardless of the domain.
Furthermore, the actual performance ratings from supervisors can be used as potential rankings of officers to reverse engineer the evaluation factors/weights to investigate potential biases.



\acknowledgments{
The authors wish to thank Steve Hawthorn, 
Patrick J. Flannelly, Christina A. Stober, and Hao Chen. 
This work is funded in part by the U.S. Department of Homeland Security VACCINE Center under Award Number 2009-ST-061-CI0003 and the Research Projects Agency-Energy (ARPA-E), U.S. Department of Energy, under Award Number DE-AR0000593.
}

\bibliographystyle{abbrv-doi-narrow}

\bibliography{metricsvis}
\end{document}